\documentclass[aps,prl,twocolumn,showpacs,groupedaddress]{revtex4}
\usepackage{amsfonts}
\usepackage{amsmath}
\usepackage{amssymb}
\usepackage{graphicx}

\begin{document}
\title{Thermally activated recovery of electrical conductivity in ${\rm LaAlO_3}/{\rm SrTiO_3}$}
\author{Snir Seri}
\author{Moty Schultz}
\author{Lior Klein}
\affiliation{Department of Physics, Nano-magnetism Research Center,
Institute of Nanotechnology and Advanced Materials, Bar-Ilan
University, Ramat-Gan 52900, Israel}


\begin{abstract}
Patterned structures of ${\rm LaAlO_3}/{\rm SrTiO_3}$ that
exhibit a decrease in their electrical conductivity below 30 K, recover their higher conductivity upon warming in a thermally activated process. Two dominant energy barriers $E_b$ are identified: $E_{b1}=0.224\pm0.003$ eV related to conductivity recovery near ${\rm 70 \ K}$  and $E_{b2}=0.44\pm0.015$ eV related to conductivity recovery near ${\rm 160 \ K}$. We discuss possible linkage to structural defects such as dislocations and twin boundaries.
\end{abstract}

\maketitle
%

An attractive feature of the interface between the insulating oxides ${\rm SrTiO_3}$  and ${\rm LaAlO_3}$ (LAO/STO) \cite{high mobility,LAO STO} is the ability to tune its transport properties by gate voltage \cite{tunable quasi,dominant mobility,superconductivity2,superconductivity3,tuning spin orbit,rashba spin orbit,CriticalDensity,phasediagram}. However, it appears that there are other mechanisms that yield effectively the same effect without gating, including similar correlations between sheet resistance, carrier density and mobility. In a recent report \cite{Rs increase}, we showed that LAO/STO patterns with current path width smaller than 10 microns may exhibit below 30 K a significant decrease in their electrical conductivity, in connection with driving a sufficiently large current through the sample and/or applying an in-plane magnetic field. The initial high conductivity is recoverable upon applying a warming cycle.

Concomitantly with the field- and current-induced decrease in conductivity, the sheet carrier density ($n_s$) and mobility decrease, magnetotransport features linked to magnetism \cite{antisymmetry,antiferromagnetism,2D_3D} are suppressed, and the nonuniformity of the sample increases. Namely, without applying a gate voltage, there are mechanisms that decrease conductivity. Furthermore, the mechanism also increases the nonuniformity of the conductivity, a feature that was directly observed with scanning probe microscopy \cite{Coexistence3,CriticalThickness}.

Here, we explore in detail the time and temperature dependence of the conductivity recovery as the sample is warmed up and show that it is well described by a thermally activated process.  We extract two energy barriers: $E_{b1}=0.224\pm0.003$ eV for the conductivity recovery near 70 K and $E_{b2}=0.44\pm0.015$ eV for the conductivity recovery near ${\rm 160 \ K}$. The conductivity exhibits a noticeable time dependence also above room temperature; however, it can not be correlated with a single thermally-activated process.

\begin{figure}[ht]
\includegraphics[scale=0.5, trim=100 0 100 0]{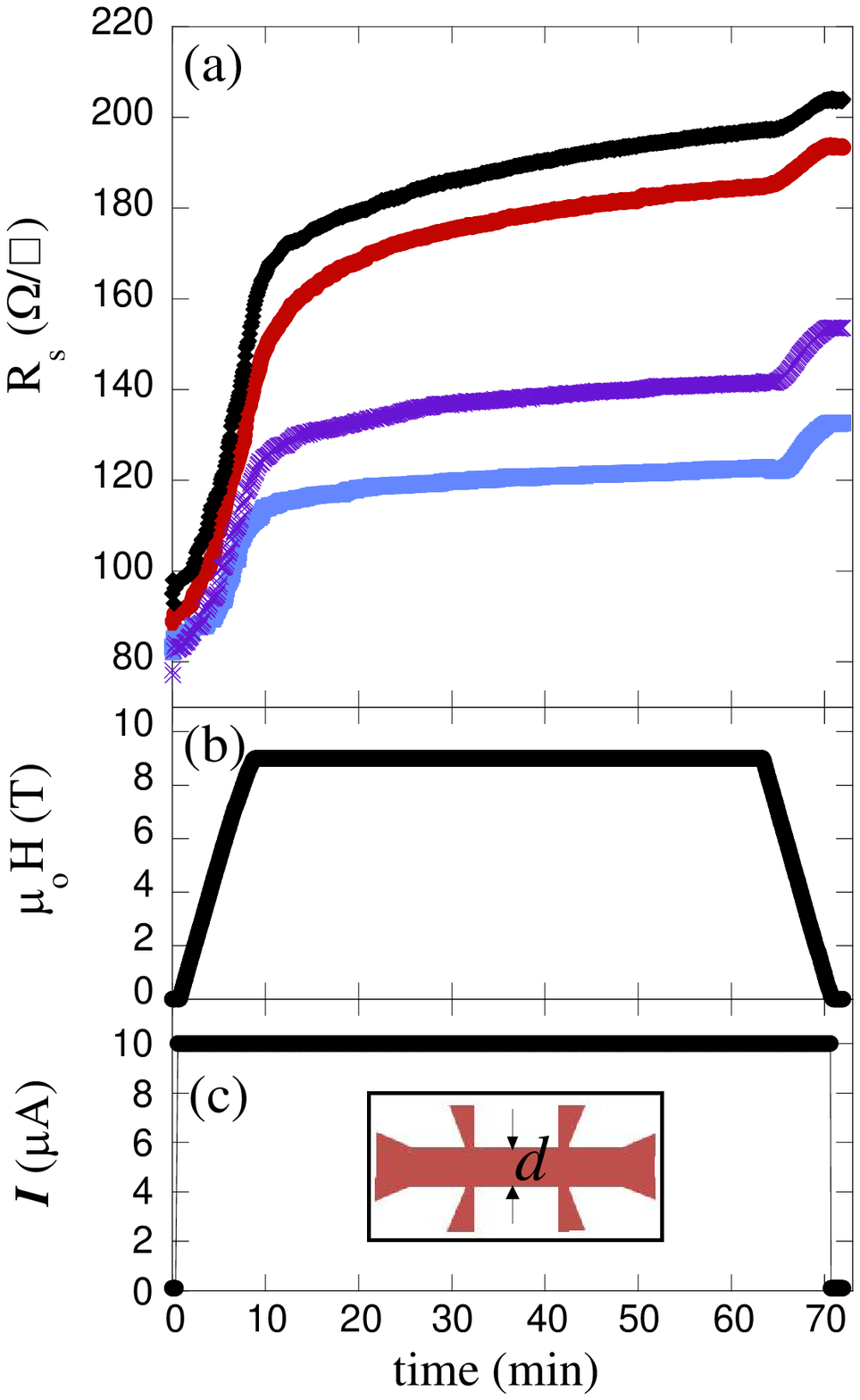}
\caption{(a) The sheet resistance ($R_s$) vs time at 5 K in a pattern with $d=10 \ \mu$m. (b) The time dependence of a magnetic field applied parallel to the LAO/STO interface. (c) The time dependence of a current driven through the pattern. Inset: A sketch of a typical pattern.}
\label{RsIncrease}
\end{figure}

\begin{figure}[ht]
\includegraphics[scale=0.5, trim=100 0 100 0]{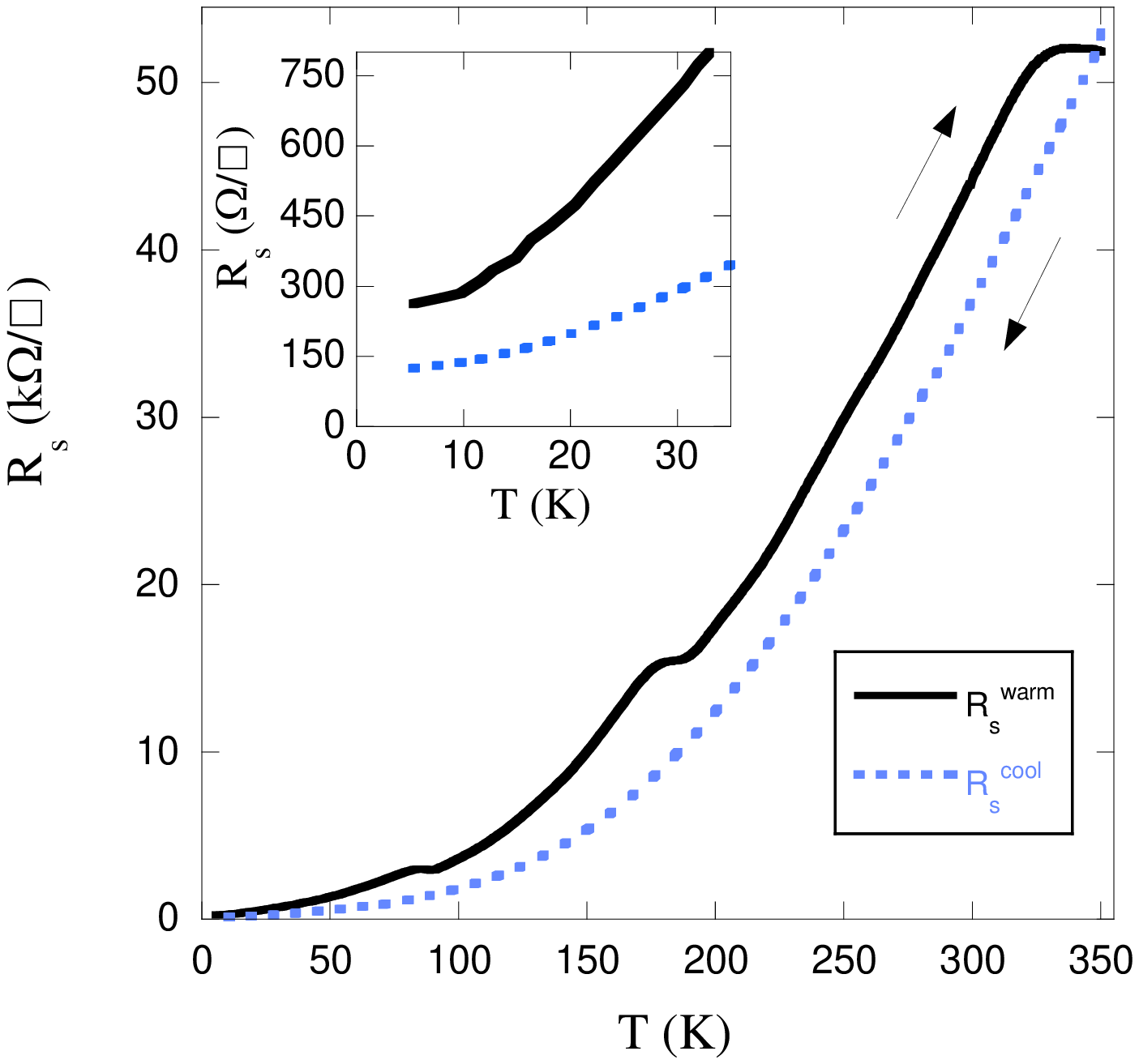}
\caption{The sheet resistance ($R_s$) as a function of temperature during cooling (dashed line, blue) and during warming (solid line, black) after increasing $R_s$ at 5 K (using the procedures shown in Fig. \ref{RsIncrease}). Inset: Blow up of the low temperature data. } \label{sweep}
\end{figure}

The results not only provide an explanation for a puzzling behavior reported previously \cite{experimental investigation,dislocations}, they also provide quantitative details on what appears to be a low-temperature charge trapping mechanism that reduces the carrier density and increases nonuniformity. Thus, they provide new insights regarding two of the main issues concerning the transport properties of the LAO/STO interface: the existence and nature of localized charge carriers \cite{localized,RIXS1,RIXS2}, and the origins of interface nonuniformity \cite{antisymmetry,Coexistence3,CriticalThickness}. The identified energy scales and length scales are instrumental in identifying the trapping sites which would enable better understanding and control of the transport properties of the LAO/STO interface. Based on the relevant length scale of the phenomenon, we suggest that the trapping sites might be linked to crystal imperfections with similar length scales, such as dislocations and twin boundaries. Interestingly, a recent report shows accelerated recovery of conductivity (after its suppression with gating) in ${\rm LaTiO_3}/{\rm SrTiO_3}$ system at temperatures close to 70 and 160 K \cite{Irreversibility LTO}, raising the possibility that the relevance of the reported phenomenon exceeds the LAO/STO interface.



%

The samples were grown by pulsed laser deposition in an oxygen atmosphere of ${\rm 7\times10^{-5}}$ mbar on ${\rm TiO_2}$ terminated (001) STO surfaces at $770^{\circ}$ C. The LAO thickness is 4 unit cells. The samples were cooled to room temperature in 400 mbars of $\rm O_2$, including one hour oxidation step at $600^{\circ}$ C. The laser fluence was about 0.8 J/cm$^2$, with repetition rate of 1 Hz. Patterning was done by photolithography as described in Ref. \cite{microlithography}. Typical geometry of our samples is shown in Fig. \ref{RsIncrease}c (inset). The current path width ($d$) in the patterns used for this research is 5 and 10 $\mu$m. The contact arrangement allows for simultaneous longitudinal and transverse voltage measurements.

%

Figure \ref{RsIncrease} shows typical protocols used to increase the sheet resistance $R_s$ at 5 K of a pattern with a current path width $d=10 \ \mu$m. The figure shows the time dependence of $R_s$ while a current of 10 $\mu$A is driven through the pattern and a magnetic field is applied parallel to the LAO/STO interface (see Figs. \ref{RsIncrease}b and \ref{RsIncrease}c). Before and after the application of the high current and the high in-plane field, $R_s$ is measured with a low current (0.1 $\mu$A) and a zero magnetic field. The different curves in Fig. \ref{RsIncrease}a are obtained with the same pattern in different cooling cycles. We note that the induced increase in $R_s$ may vary significantly in different cooling cycles.

\begin{figure}[ht]
\includegraphics[scale=0.45, trim=100 0 100 0]{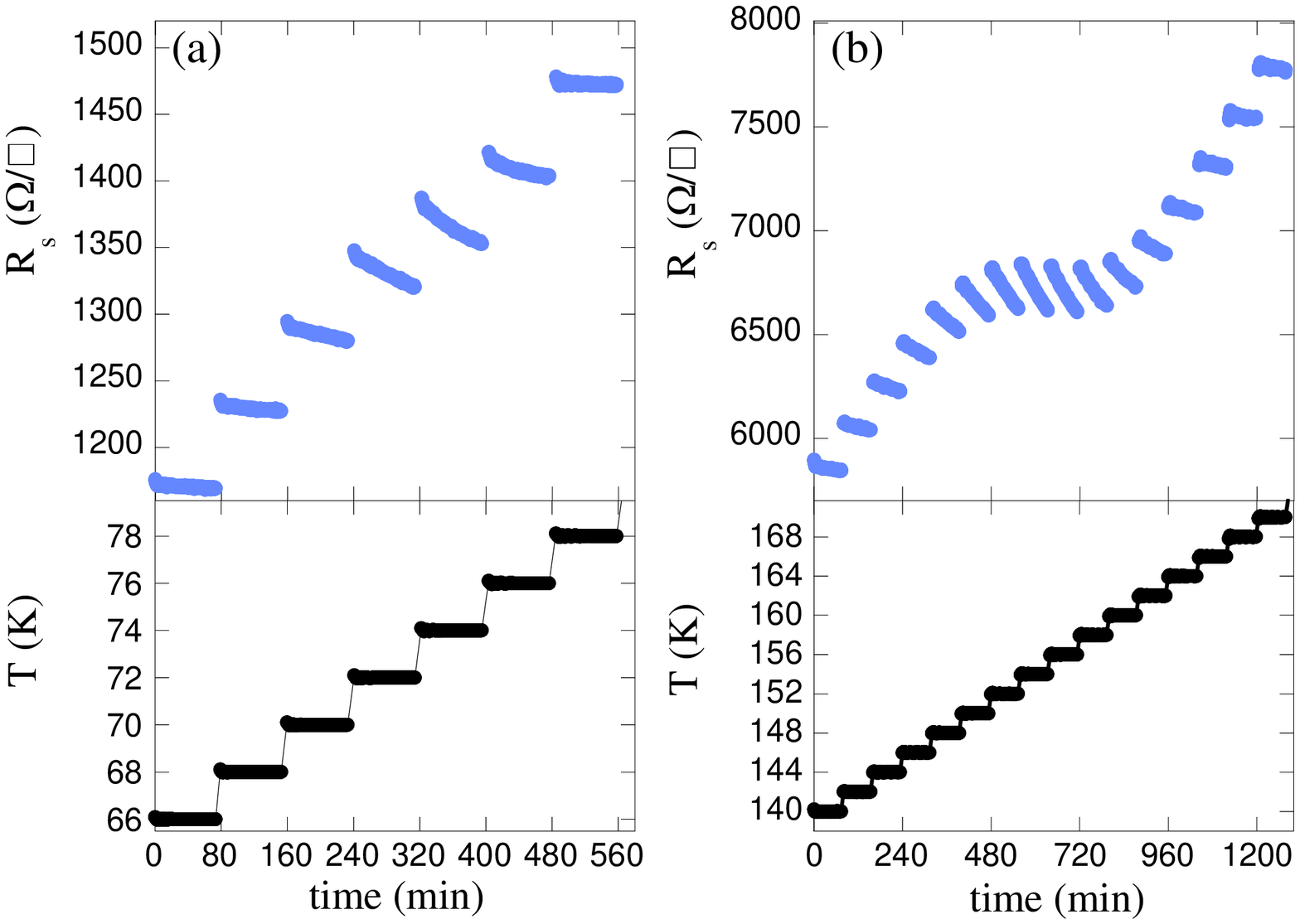}
\caption{The sheet resistance ($R_s$) as a function of time after increasing $R_s$ at 5 K (see Fig. \ref{RsIncrease}), as the temperature is increased in steps near 70 K (a) and 160 K (b).} \label{Release}
\end{figure}

\begin{figure}[ht]
\includegraphics[scale=0.47, trim=100 0 100 0]{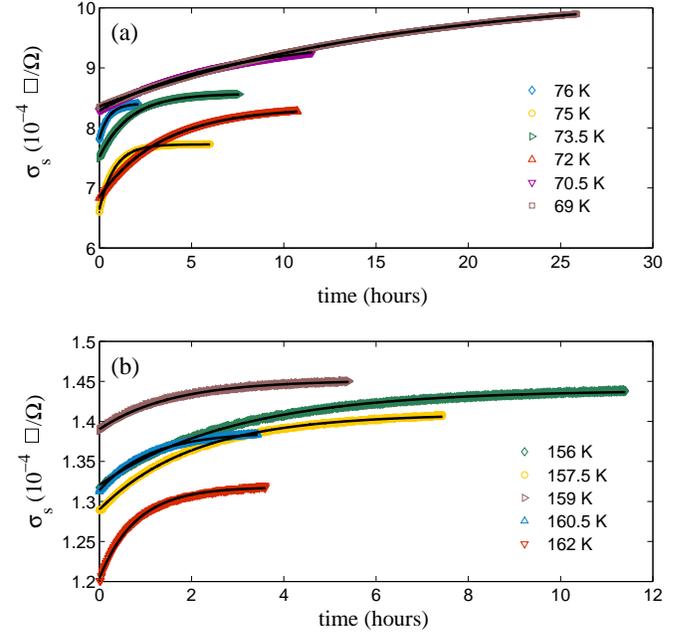}
\caption{The sheet conductivity ($\sigma_s$) as a function of time after increasing $R_s$ at 5 K (see Fig. \ref{RsIncrease}), at different temperatures near 70 K (a) and 160 K (b). After each measurement, $R_s$ was recovered to its as-cooled value by warming the sample to 350 K. The lines are fits to Eq. \ref{sigma}.} \label{Longtime}
\end{figure}

Figure \ref{sweep} shows the temperature dependence of $R_s$ during cooling (dashed line, blue) and warming (solid line, black) after applying at 5 K the protocols to increase $R_s$ as shown in Figure \ref{RsIncrease}. The temperature is changed continuously at a rate of about 8 K/min both in cooling and warming. The breaks in the warming curve near 70 and 160 K suggest an accelerated decrease of $R_s$. To understand its nature, we focus on the time dependence of $R_s$ in the vicinity of the two temperatures.

\begin{figure}[ht]
\includegraphics[scale=0.5, trim=100 0 100 0]{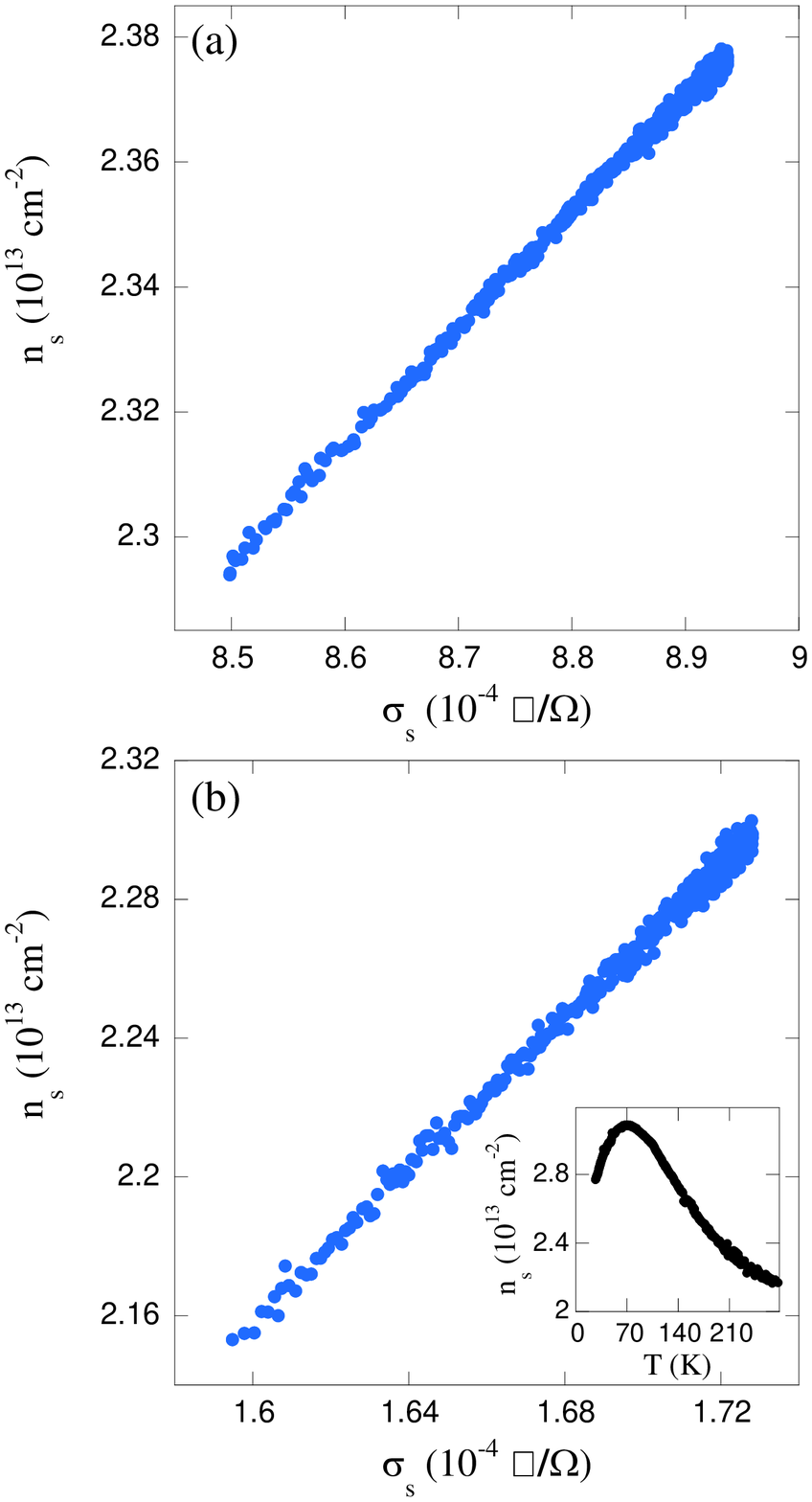}
\caption{The sheet carrier density ($n_s$) as a function of the sheet conductivity ($\sigma_s$) at 75 K (a) and 160.5 K (b). Inset: The temperature dependence of $n_s$ in the relevant temperature range during cooling.} \label{ns}
\end{figure}

Figures \ref{Release}a and \ref{Release}b show $R_s$ as a function of time (after a low-temperature increase in $R_s$) as the temperature is increased in steps near 70 and 160 K, respectively. For the two temperature intervals, the recovery rate increases with increasing temperature until it saturates.

Figures \ref{Longtime}a and \ref{Longtime}b demonstrate the recovery in a different way. They show the time dependence of the sheet conductivity $\sigma_s$ at different temperatures near 70 and 160 K where each measurement is performed after warming the sample to 350 K, cooling it to 5 K, and increasing its sheet resistance as shown in Fig. \ref{RsIncrease}. The same protocol was used for all measurements. Nevertheless, non-monotonic behavior of $\sigma_s$ as a function of temperature is observed in Figs. \ref{Longtime}a and \ref{Longtime}b due to the different induced $R_s$ increases obtained in the different cooling cycles, as shown in Fig. \ref{RsIncrease}.

\begin{figure}[ht]
\includegraphics[scale=0.45, trim=100 0 100 0]{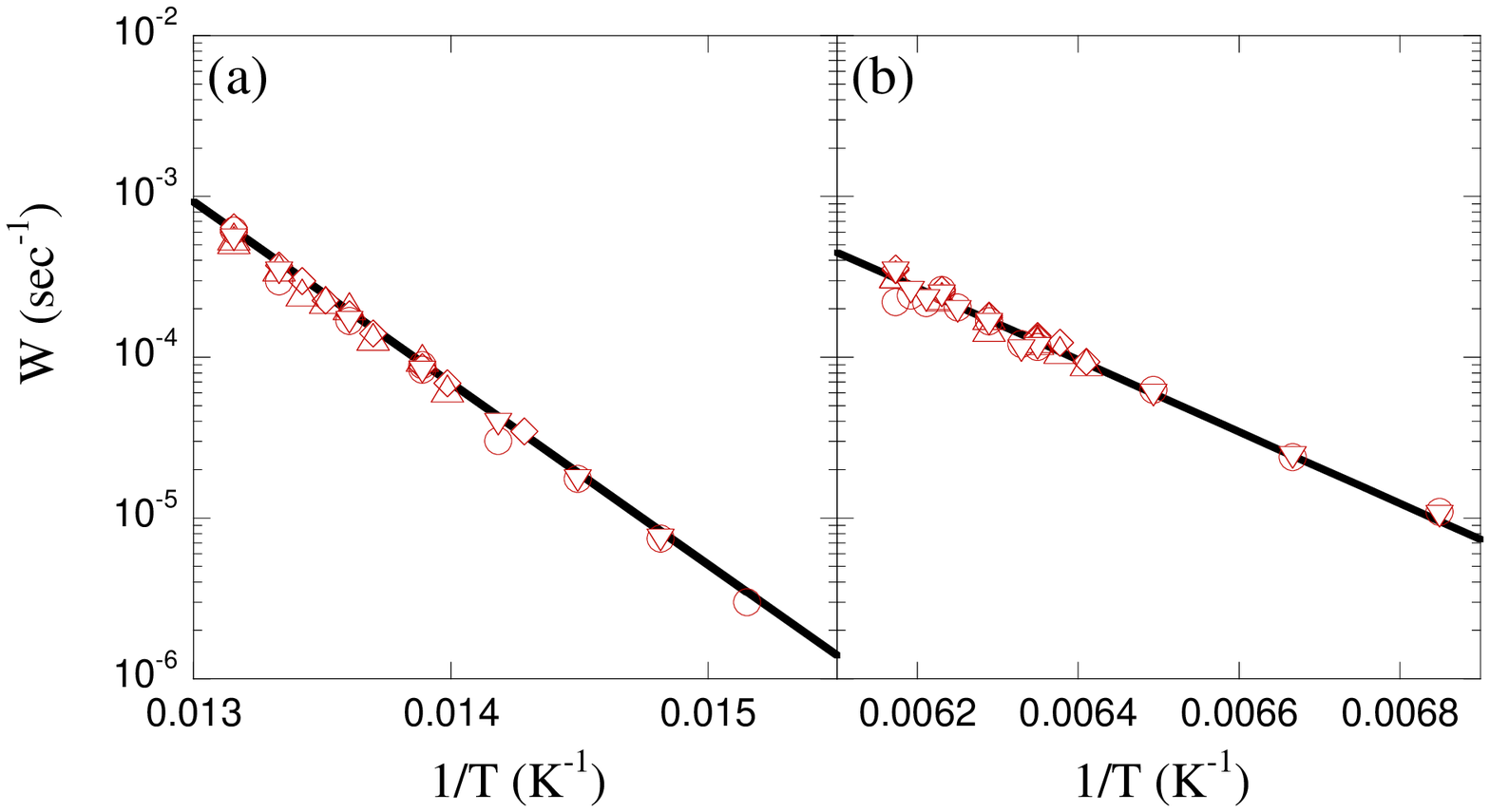}
\caption{The parameter $W$ (Eq. \ref{sigma}) as a function of 1/\emph{T} near 70 K (a) and 160 K (b). Different symbols represent different patterns. The lines are fits to Eq. \ref{sigma2}. } \label{W12}
\end{figure}

Figures \ref{ns}a and \ref{ns}b show the change in the sheet carrier density $n_s$, extracted from Hall effect measurements, as a function of $\sigma_s$ as it recovers with time at 75 and 160.5 K, respectively. The inset of Figure \ref{ns}b shows the temperature dependence of the extracted $n_s$ in the relevant temperature range during cooling the sample, when no relaxation effects exist. The Hall effect resistance was extracted by exchanging the current and voltage leads without reversing the field, as described in Ref. \cite{antisymmetry}. For the Hall measurements we apply perpendicular fields up to 9 T with no observable effect on the rate of the conductivity recovery. We note that there is a linear correlation between $n_s$ and $\sigma_s$ during the conductivity recovery; therefore, to describe the conductivity recovery we may use a model for the time dependence of $n_s$.

We assume that the low-temperature increase in $R_s$ is due to induced trapping of charge carriers and that the high-temperature recovery is due to their thermally-activated release. The corresponding rate equations are
\begin{equation}
\frac{dN_1}{dt}=-W_{12}{\cdot}N_1+W_{21}{\cdot}N_2
\label{dN1}
\end{equation}
\begin{equation}
\frac{dN_2}{dt}=W_{12}{\cdot}N_1-W_{21}{\cdot}N_2
\label{dN2}
\end{equation}
where $N_1$ and $N_2$ are the number of trapped and untrapped charge carriers, respectively, and $W_{12}$ ($W_{21}$) is the probability for a trapped (untrapped) charge carrier to be released (trapped).

Considering the linear dependence between $\sigma$ and $n_s$ during the conductivity recovery, we obtain
\begin{equation}
\frac{d\sigma}{dt} \propto (N^0_1{\cdot}W_{12}-N^0_2{\cdot}W_{21}){\cdot}e^{-W {\cdot} t}
\label{dsigma1}
\end{equation}
where $N^0_1$ ($N^0_2$) is the initial number of the trapped (untrapped) charge carriers and $W=W_{12}+W_{21}$.


From here we find that
\begin{equation}
\sigma(t)=-A {\cdot} e^{-W {\cdot} t}+B
\label{sigma}
\end{equation}
where the coefficient B is the conductivity at $t\rightarrow\infty$ and A=$\sigma_{t\rightarrow\infty} - \sigma_{t=0}$. The lines in Figures \ref{Longtime}a and \ref{Longtime}b are fits to Eq. \ref{sigma}.

Figures \ref{W12}a and \ref{W12}b show the parameter $W$ [Eq. \ref{sigma}] in a logarithmic scale as a function of 1/\emph{T} near 70 and 160 K, respectively, extracted by analyzing the time-dependent conductivity of several patterns of two different samples. The clear linear dependence indicates
\begin{equation}
W=f_0 \cdot e^{-E_b/k_BT}
\label{sigma2}
\end{equation}
suggesting an Arrhenius-type behavior. The lines in Figs. \ref{W12}a and \ref{W12}b are fits to Eq. \ref{sigma2}.

We identify two dominant energy barriers: $E_{b1}=0.224\pm0.003$ eV related to conductivity recovery near ${\rm 70 \ K}$  and $E_{b2}=0.44\pm0.015$ eV related to conductivity recovery near ${\rm 160 \ K}$. The value of $f_0$ is on the order of $10^{11}$ ${\rm s^{-1} }$ for the conductivity recovery near ${\rm 70 \ K}$ and on the order of $10^{10}$ ${\rm s^{-1} }$ for the conductivity recovery near ${\rm 160 \ K}$.

The time dependence of the conductivity above room temperature is more complicated and pattern dependent. In several patterns, the conductivity increases initially; however, after some time it starts decreasing and appears to saturate. This may indicate that the time dependence of the conductivity  above room temperature is affected by several processes. We note that assuming two competing relaxation processes yields
\begin{equation}
\sigma(t)=-A {\cdot} e^{-W {\cdot} t}+A' {\cdot} e^{-W' {\cdot} t}+B.
\label{modifiedsigma}
\end{equation}
which fits the data quite well. However, the fitting parameters are strongly pattern dependent so no clear conclusion can be obtained.

Based on our measurements, a plausible scenario is that the current- and field-induced suppression of conductivity  below 30 K is due to trapping of charge carriers in sites characterized by well-defined trapping energies of $E_{b1}=0.224\pm0.003$ eV and $E_{b2}=0.44\pm0.015$ eV. As trapping occurs only below 30 K, the existence of trapping sites does not affect conductivity in cooling; however, they do lead to the observed breaks in resistivity upon warming \cite{experimental investigation,dislocations},
provided low-temperature charge trapping occurred. The two trapping energies are responsible for the recovery in the vicinity of 70 and 160 K. Some conductivity recovery takes place also above room temperature; however, we could not extract specific trapping energies responsible for the recovery, probably due to the fact that more than one process takes place simultaneously.

We can not identify based on our results what are the trapping sites. A very significant hint, however, is the characteristic length scale. As we noted \cite{Rs increase}, the low-temperature current- and field-induced conductivity suppression becomes significant in patterns with length scale on the order of microns. Furthermore, in patterns with length scale on the order of 2 microns the conductivity is occasionally suppressed by orders of magnitude. In addition, the conductivity suppression is accompanied by increased spatial variation in conductivity. Therefore, it appears that the trapping sites are related to crystal imperfections of similar length scales. Possible candidates are dislocations that were found to reduce conductivity and mobility of the LAO/STO interface with unusual temperature dependence during warming which is reminiscent to our results \cite{dislocations}, or twin boundaries in ${\rm SrTiO_3}$ which form due to structural phase transitions of the STO \cite{110K,structural phase transition,Dielectric Properties}.

L.K. acknowledges support by the German Israeli Foundation (Grant
No. 979/2007) and by the Israel Science Foundation founded by the
Israel Academy of Sciences and Humanities (Grant No. 577/07). We
acknowledge the contribution of Jochen Mannhart and Rainer Jany who provided the samples used for this research.


\begin{thebibliography}{9}

\bibitem{high mobility} A. Ohtomo and H. Y. Hwang, Nature \textbf{427}, 423 (2004).

\bibitem{LAO STO} M. Huijben, A. Brinkman, G. Koster, G. Rijnders, H. Hilgenkamp, and D. H. A. Blank, Adv. Mater. \textbf{21}, 1665 (2009).

\bibitem{tunable quasi} S. Thiel, G. Hammerl, A. Schmehl, C. W. Schneider, and J. Mannhart, Science \textbf{313}, 1942 (2006).

\bibitem{dominant mobility} C. Bell, S. Harashima, Y. Kozuka, M. Kim, B. G. Kim, Y. Hikita, and H.Y. Hwang, Phys. Rev. Lett. \textbf{103}, 226802 (2009).

\bibitem{superconductivity2} A. D. Caviglia, S. Gariglio, N. Reyren, D. Jaccard, T. Schneider, M. Gabay, S. Thiel, G. Hammerl, J. Mannhart, and J. -M. Triscone, Nature \textbf{456}, 624 (2008).

\bibitem{superconductivity3} T. Schneider, A. D. Caviglia, S. Gariglio, N. Reyren, and J. -M. Triscone, Phys. Rev. B \textbf{79}, 184502 (2009).

\bibitem{tuning spin orbit} M. Ben Shalom, M. Sachs, D. Rakhmilevitch, A. Palevski, and Y. Dagan, Phys. Rev. Lett. \textbf{104}, 126802 (2010).

\bibitem{rashba spin orbit} A. D. Caviglia, M. Gabay, S. Gariglio, N. Reyren, C. Cancellieri, and J. -M. Triscone, Phys. Rev. Lett. \textbf{104}, 126803 (2010).

%
%



\bibitem{CriticalDensity} A. Joshua, S. Pecker, J. Ruhman, E. Altman and S. Ilani, Nat. Commun. \textbf{3}, 1129 (2012).

\bibitem{phasediagram} A. Joshua, J. Ruhman, S. Pecker, E. Altman and S. Ilani, arXiv:1207.7220.



%

\bibitem{Rs increase} S. Seri, M. Schultz, and L. Klein, Phys. Rev. B \textbf{86}, 085118 (2012).

\bibitem{antisymmetry} S. Seri and L. Klein, Phys. Rev. B \textbf{80}, 180410(R) (2009).

\bibitem{antiferromagnetism} M. Ben Shalom, E. Levy, A. Palevski, Y. Dagan, C. W. Tai, and Y. Lereah, Phys. Rev. B \textbf{80}, 140403(R) (2009).

\bibitem{2D_3D} X. Wang, W. M. L$\ddot{u}$, A. Annadi, Z. Q. Liu, K. Gopinadhan, S. Dhar, T. Venkatesan, and Ariando, Phys. Rev. B \textbf{84}, 075312 (2011).

%
%

\bibitem{Coexistence3} J. A. Bert, B. Kalisky, C. Bell, M. Kim, Y. Hikita, H. Y. Hwang, and K. A. Moler, Nature Phys. \textbf{7}, 767 (2011).

\bibitem{CriticalThickness} B. Kalisky, J. A. Bert, B. B. Klopfer, C. Bell, H. K. Sato, M Hosoda, Y. Hikita, H. Y. Hwang, and K. A. Moler, Nat. Commun. \textbf{3}, 922 (2012).

\bibitem{experimental investigation} W. Siemons, G. Koster, H. Yamamoto, T. H. Geballe, D. H. A. Blank, and M. R. Beasley, Phys. Rev. B \textbf{76}, 155111 (2007).

\bibitem{dislocations} S. Thiel, C. W. Schneider, L. Fitting Kourkoutis, D. A. Muller, N. Reyren, A. D. Caviglia, S. Gariglio, J.-M. Triscone, and J. Mannhart, Phys. Rev. Lett. \textbf{102}, 046809 (2009).

\bibitem{localized} Z. S. Popovi\'{c}, S. Satpathy, and R. M. Martin, Phys. Rev. Lett. \textbf{101}, 256801 (2008).

\bibitem{RIXS1} G. Berner, S. Glawion, J. Walde, F. Pfaff, H. Hollmark, L.-C. Duda, S. Paetel, C. Richter, J. Mannhart, M. Sing, and R. Claessen, Phys. Rev. B \textbf{82}, 241405(R) (2010).

\bibitem{RIXS2} K. Zhou, M. Radovic, J. Schlappa, V. Strocov, R. Frison, J. Mesot, L. Patthey, and T. Schmitt, Phys. Rev. B \textbf{83}, 201402(R) (2011).

\bibitem{Irreversibility LTO} J. Biscaras, S. Hurand, C. Feuillet-Palma, A. Rastogi, R. C. Budhani, N. Reyren, E. Lesne, D. LeBoeuf, C. Proust, J. Lesueur, and N. Bergeal, arXiv:1206.1198.

%





\bibitem{microlithography} C. W. Schneider, S. Thiel, G. Hammerl, C. Richter, and J. Mannhart, Appl. Phys. Lett. \textbf{89}, 122101 (2006).



\bibitem{110K} L. Rimai and G. deMars, Phys. Rev. \textbf{127}, 702 (1962).

\bibitem{structural phase transition} Farrel W. Lytle, J. Appl. Phys. \textbf{35}, 2212 (1964).

\bibitem{Dielectric Properties} T. Sakudo and H. Unoki, Phys. Rev. Lett. \textbf{26}, 851 (1971).

%
%
%
%
%


\end{thebibliography}
\end{document}